\documentclass[12pt]{article}

\usepackage{graphicx}
\usepackage{amssymb}
\usepackage{amsmath}
\usepackage{color}

\def\bea{\begin{eqnarray}}
\def\eea{\end{eqnarray}}
\def\beq{\begin{equation}}
\def\eeq{\end{equation}}

\def\bm{\begin{math}}
\def\me{\end{math}}

\begin{document}

\begin{center}
{\Large{\bf Domain Growth in Chiral Phase Transitions: Role of Inertial Dynamics}} \\
\ \\
\ \\
by \\
Awaneesh Singh$^1$, Sanjay Puri$^1$ and Hiranmaya Mishra$^{1,2}$ \\
$^1$School of Physical Sciences, Jawaharlal Nehru University, New Delhi--110067, India. \\
$^2$Theory Division, Physical Research Laboratory, Navrangpura, Ahmedabad--380009, India.
\end{center}

\begin{abstract}
We investigate the kinetics of phase transitions for chiral symmetry breaking in heavy-ion collisions. We use a Langevin description for order-parameter kinetics in the chiral transition. The Langevin equation of motion includes {\it dissipation} and an {\it inertial term}. We study the ordering dynamics subsequent to a quench from the massless quark phase to the massive quark phase, and discuss the effect of inertia on the growth kinetics.
\end{abstract}

\newpage
\section{Introduction}
\label{intro}

Strongly-interacting hadronic matter is expected to undergo a phase transition at sufficiently high temperature and baryon density \cite{rischke}. For vanishing baryon density, this is a prediction from {\it ab initio} calculations like lattice simulations of {\it quantum chromodynamics} (QCD) \cite{karsch}. For finite baryon density, this is expected from different effective models of strongly-interacting systems. Heavy ion collisions provide experimental opportunities to study QCD transitions. The high temperature and low baryon density phase was explored in the {\it relativistic heavy ion collision} (RHIC) experiments at Brookhaven, and will be further studied in {\it large hadron collider} (LHC) experiments. Some planned experiments like {\it beam energy scan} at RHIC \cite{rhic}; {\it compressed baryonic matter} at GSI \cite{cbm}; and the {\it nuclotron-based ion collider facility} (NICA) at Dubna \cite{nica} intend to study different aspects of the thermodynamic properties of QCD at finite chemical potential, e.g., the expected critical point of QCD, first-order phase transitions, mixed phase structures, etc.

It is challenging to extract the thermodynamic properties of quark-hadron phase transitions from nuclear collision experiments due to the absence of global thermal equilibrium. This is because nonequilibrium effects play an important role in the evolution of the fireball. Therefore, one also has to understand the kinetic processes that drive the phase transitions, and the properties of the nonequilibrium structures that the system forms on its way to equilibrium \cite{aj94,pw09}. In this context, both {\it critical dynamics} and the {\it far-from-equilibrium kinetics} of chiral transitions have attracted recent attention. In the study of critical dynamics (i.e., the time-dependent behavior in the vicinity of the critical point), much interest has focused upon the signatures of the {\it critical end point} (CEP) of QCD \cite{berdinkov,son,koide,tomoinpa}.

In this paper and an earlier companion paper \cite{ak10}, we focus on far-from-equilibrium kinetics, i.e., the evolution of the system after a quench from a disordered phase to an ordered phase with non-vanishing quark-anti-quark condensates. In this context, the relaxation to equilibrium in a Langevin framework has been studied by Fraga and Krein \cite{fragaplb}. These authors studied the early-time dynamics of {\it spinodal decomposition} (i.e., spontaneous kinetics) and the effect of dissipation on the spinodal instability. Further, the bubble {\it nucleation kinetics} in chiral transitions was studied by Bessa et al. \cite{fraga}. A {\it time-dependent Ginzburg-Landau} (TDGL) equation for this problem was derived in Ref.~\cite{skokovdima}, starting from a non-ideal, non-relativistic hydrodynamics for coupled order parameters. Further, Randrup \cite{jr09} studied the amplification of spinodal fluctuations within a fluid-dynamical model for the nuclear collisions. Randrup's study focused on the evolution in the linearized regime, which showed an exponential growth of the initial fluctuations.

We recently initiated a study of far-from-equilibrium kinetics of chiral phase transitions \cite{ak10}. Our approach was complementary to the studies of Refs.~\cite{fragaplb,fraga,skokovdima,jr09}: we used a TDGL equation to investigate the late stages of phase-separation kinetics in quark matter. The Ginzburg-Landau (GL) free-energy functional was obtained from a Taylor expansion of the thermodynamic potential for a two-flavor {\it Nambu-Jona-Lasinio} (NJL) model. We studied domain growth subsequent to quenches through the first-order and second-order lines of the phase diagram. We examined the quantitative features of the coarsening morphologies in both cases.

Let us note that the TDGL equation, which models the overdamped (relaxational) dynamics of an order-parameter field to the minimum of the thermodynamic potential \cite{hh77}, is first-order in the time-derivative. The inertial term with a second-order time-derivative is usually neglected in comparison to the damping term. However, a microscopic derivation of the equation in a relativistic field theory using, e.g., the {\it closed-time-path Green's function}  (CTPGF) formalism leads to a second-order stochastic equation. Such a derivation has been done for scalar field theories \cite{bd99,gr94,dr98}. A second-order TDGL equation has also been derived for the NJL model in Ref.~\cite{fh11} using the CTPGF method. More recently, a Langevin equation with an inertial term has been derived for the chiral order parameter field in a sigma model including quark degrees of freedom by Nahrgang et al. \cite{nl11}. These authors use an influence-functional method and calculate the explicit form of the damping coefficient, as well as the form of noise correlators that appear in the Langevin equation. This model has been used to discuss the relaxational dynamics of the order parameter near the critical point \cite{nahrgang2,nahrgang3}.

Given this background, it is very relevant to investigate the effect of an inertial term on the ordering kinetics of the chiral transition. More generally, it is important to study the effect of an inertial term in domain growth problems. In spite of the intense interest in the kinetics of phase transitions, this question has received almost no attention \cite{aj94,pw09}. In this paper, we will address this issue in the context of chiral transitions.

This paper is organized as follows. In Sec.~\ref{MM}, we recapitulate the NJL model and its mapping to the $M^6$-Landau potential discussed in Ref. \cite{ak10}. In Sec.~\ref{DE}, this is used to formulate a Langevin equation for the order parameter evolution with an inertial term. In Secs.~\ref{case1} and \ref{case2}, we study the kinetics of chiral phase transitions resulting from different quenches. Our primary interest is the effect of the inertial term on the ordering dynamics. Finally, Sec.~\ref{summary} concludes this paper with a summary and discussion.

\section{Thermodynamic Potential and Phase Diagram}
\label{MM}

To discuss chiral phase transitions, we model chiral symmetry breaking in strong interactions by the 2-flavor NJL model. The thermodynamic potential in terms of constituent mass is given as \cite{ak10}
\begin{align}
\tilde\Omega(M,\beta,\mu) =& \;-\dfrac{12}{(2\pi)^3\beta}\displaystyle\int \! d\vec{k}\;\Big\{ \ln\left[1+e^{-\beta\left( \sqrt{k^2+M^2} - \mu \right)}\right] \notag\\
& \qquad\qquad\qquad\quad + \ln\left[1+e^{-\beta\left( \sqrt{k^2+M^2} + \mu \right)}\right] \Big\}   \notag\\
&\; -\dfrac{12}{(2\pi)^3}\displaystyle\int \! d\vec{k} \; \left(\sqrt{k^2+M^2}-k\right)+ \dfrac{M^2}{4G} ,
\label{tomega}
\end{align}
where $\beta = (k_B T)^{-1}$ and $\mu$ is the chemical potential. In Eq.~(\ref{tomega}), we have taken vanishing current quark mass $m=0$. We introduce $M=-2g\rho_s$, with $\rho_s=\langle\bar\psi\psi\rangle$ being the scalar density; and $g=G[1+1/(4N_c)]$, $N_c$ being the number of colors. We have taken the four-fermion coupling $G=5.0163\times 10^{-6}$ $\mathrm{MeV^{-2}}$, {and the three-momentum ultraviolet cut-off $\Lambda=653.3$ MeV} \cite{ay89}. 

Close to the phase boundary, the potential in Eq.~(\ref{tomega}) may be expanded as a Landau potential in the order parameter $M$ \cite{ak10}: 
\begin{equation}
\tilde\Omega\left(M \right)= \tilde\Omega\left(0 \right) + \frac{a}{2}M^2 + \frac{b}{4}M^4 + \frac{d}{6}M^6 + O(M^8) \equiv f(M),
\label{p6}
\end{equation}
correct up to logarithmic factors \cite{cs08}. In the following, we consider the expansion of $\tilde\Omega\left(M \right)$ up to the $M^6$-term. This will prove adequate to recover the phase diagram of the NJL model \cite{ak10}. 
The first two coefficients in Eq.~(\ref{p6}) can be obtained by comparison with Eq.~(\ref{tomega}) as
\begin{align}
\tilde\Omega(0) =&\;-\dfrac{6}{\pi^2\beta}\displaystyle\int_0^\Lambda \!\!\! dk\,\, k^2 \left\lbrace  
\ln\left[1+e^{-\beta(k-\mu)}\right] + \ln\left[1+e^{-\beta(k+\mu)}\right]\right\rbrace, \nonumber \\
a =& \; \dfrac{1}{2G} - \dfrac{3\Lambda^2}{\pi^2} + \dfrac{6}{\pi^2}\displaystyle\int_0^\Lambda \!\!\! dk\,\,k\left[ \dfrac{1}{1+e^{\beta(k-\mu)}} + \dfrac{1}{1+e^{\beta(k+\mu)}}\right]. 
\label{coff}
\end{align}
{We treat the higher coefficients as phenomenological parameters, which are obtained by fitting $\tilde\Omega\left(M \right)$ in Eq.~(\ref{p6}) to the integral expression for $\tilde\Omega (M)$ in Eq.~(\ref{tomega}). There are two free parameters in the microscopic theory ($T$ and $\mu$), so we consider the $M^6$-Landau potential with parameters $b$ and $d$.} For stability, we require $d>0$.

\begin{figure}[!htb]
\centering
\includegraphics[width=1.0\textwidth]{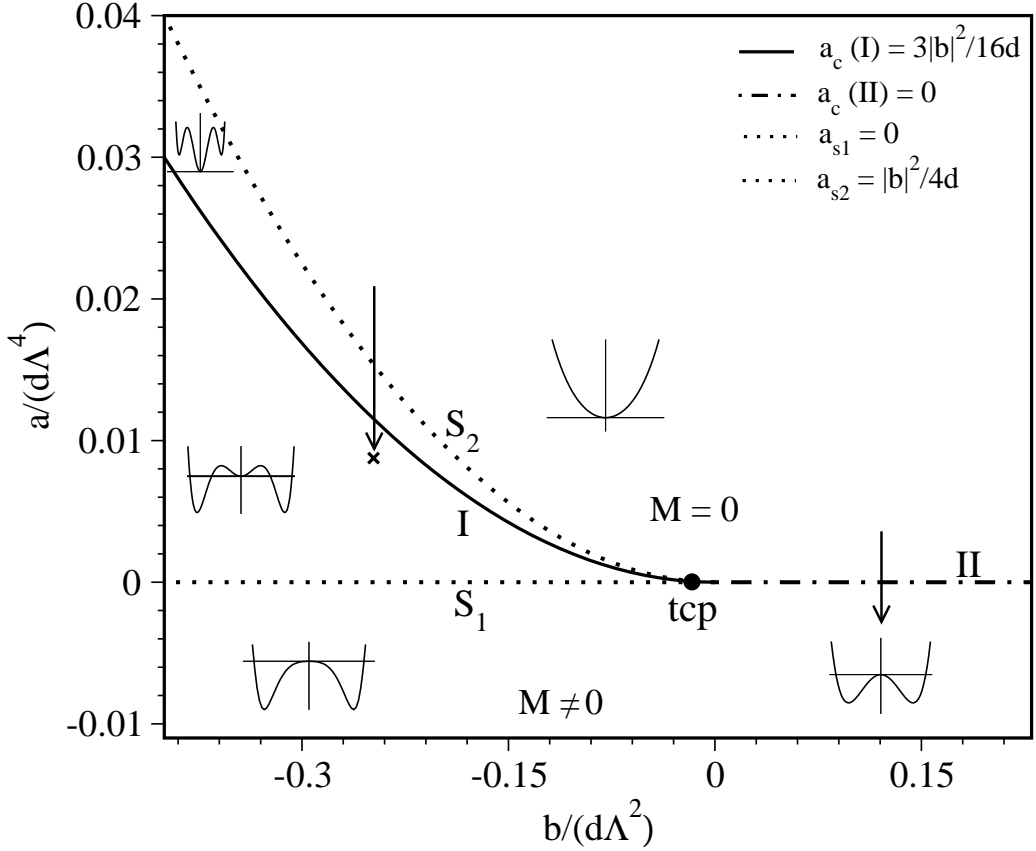}
\caption{Phase diagram for the Landau free energy in Eq.~(\ref{p6}) in the [$b/(d\Lambda^2), a/(d\Lambda^4)$]-plane. A line of first-order transitions (I) meets a line of second-order transitions (II) at the tricritical point (tcp), which is located at $a=b=0$. The equations for I and II are specified in the figure. The dotted lines denote the spinodals $S_1$ and $S_2$, whose equations are also provided. The typical forms of the Landau potential in various regions are shown in the figure.  {The cross  denotes the point where we quench the system for $b<0$ (first-order quench). The second-order quench considered in Sec.~\ref{case1} corresponds to $b/(d\Lambda^2) = 1.269$, $a/(d\Lambda^4) = -0.225$, and is not shown in the figure for clarity.}}
\label{fig1}
\end{figure}

The gap equation, $f'\left(M \right)= aM + bM^3 + dM^5=0$, yields five solutions for the order parameter as $M_0=0$ and $M_{\pm}^2=(-b\pm \sqrt{b^2 - 4ad})/(2d)$. In Fig.~\ref{fig1} we show the phase diagram for the Landau potential in $[b/(d\Lambda^2), a/(d\Lambda^4)]$-space \cite{ak10}. For $b>0$, the transition is second-order. The stationary points are (i) $M=0$ (for $a>0$), and (ii) $M=0$, $\pm M_+$ (for $a<0$). For $a<0$, the preferred equilibrium state is the one with massive quarks \cite{mi04}. Next, for $b<0$, the corresponding solutions of the gap equation are (i) $M=0$ [for $a>b^2/(4d)$], (ii) $M =0,\,\pm{M_+},\, \pm{M_-}$ [for $b^2/(4d)>a>0$], and (iii) $M=0,\,\pm{M_+}$ (for $a<0$). As we reduce $a$ from higher values, 5 solutions appear at $a=|b|^2/(4d)$. However, this does not correspond to a phase transition. On further reduction of $a$, a first-order phase transition occurs at $a_c=3|b|^2/(16d)$. The order parameter jumps discontinuously from $M=0$ to  $M=\pm M_+$, where $M_+ = [3|b|/(4d)]^{1/2}$. The first-order line meets the second-order line in a tricritical point, which is located at $b_\text{tcp}=0$, $a_\text{tcp}=0$. {(This corresponds to the temperature $T_{\rm tcp} = 78$ MeV and chemical potential $\mu_{\rm tcp} = 282.58$ MeV \cite{ak10}.)} The dotted lines in Fig.~\ref{fig1} denote the spinodals $S_1$ and $S_2$, with equations $a_{S_1}=0$ and $a_{S_2} = |b|^2/(4d)$. The typical forms of the Landau potential in various regions are shown in Fig.~\ref{fig1}. The cross denotes the point where we quench the system for $b<0$ (discussed in Sec.~\ref{case2}).

\section{The Dynamical Equation}
\label{DE}

We investigate the time-dependent behavior of the order parameter {$M(\vec{r},t)$}, and its approach to equilibrium, within the framework of Langevin dynamics. The evolution of the system is described by a Langevin equation with an inertial term:
\beq
\frac{\partial^2}{\partial t^2} M (\vec{r}, t)+ \bar{\gamma} \frac{\partial M}{\partial t} = -\frac{\delta \Omega\left[M \right]}{ \delta M(\vec r,t)} + \theta\left(\vec{r},t\right) ,
\label{ke1}
\eeq
where $\bar{\gamma}$ is the dissipation coefficient. {(The motivation for considering this inertial TDGL equation has been provided in Sec.~\ref{intro}.)} The coarse-grained free-energy functional $\Omega\left[M\right]$ depends on the order parameter field $M(\vec{r},t)$ as follows:
\begin{eqnarray}
\Omega\left[M\right] &=& \int d\vec{r} \left[f(M) + \frac{K}{2} \left(\vec{\nabla}M\right)^2\right] \nonumber \\
&=& \int \! d\vec{r} \left[\frac{a}{2}M^2 + \frac{b}{4}M^4 + \frac{d}{6}M^6 + \frac{K}{2} \left(\vec{\nabla}M\right)^2\right].
\label{Om}
\end{eqnarray}
Here $K$ measures the energy cost of spatial inhomogeneities. {The thermal noise $\theta(\vec{r},t)$ is stochastic, and is assumed to be Gaussian and white, satisfying the fluctuation-dissipation theorem \cite{hh77}}
\bea
\left\langle \theta\left(\vec{r},t\right) \right\rangle &=& 0 , \nonumber \\
\left\langle \theta(\vec{r'},t')\theta(\vec{r''},t'') \right\rangle &=& 2 \bar{\gamma} T\delta(\vec{r'}-\vec{r''})\delta\left(t'-t''\right) .
\label{fdr0}
\eea

We substitute Eq.~(\ref{Om}) in Eq.~(\ref{ke1}), and introduce dimensionless variables:
\begin{eqnarray}
M &=& M_0 M', \quad M_0=\sqrt{|a|/|b|}, \nonumber \\
\vec{r} &=& \xi \vec{r'}, \quad \xi=\sqrt{K/|a|}, \nonumber \\
t &=& t_0 t', \quad t_0=1/\sqrt{|a|}, \nonumber \\
\theta &=& |a|M_0~\theta' .
\label{scale}
\end{eqnarray}
Dropping primes, we then have the dimensionless form of the evolution equation:
\begin{align}
\frac{\partial^2 M}{\partial t^2} + \gamma\frac{\partial M}{\partial t} = -\mathrm{sgn} (a) M -\mathrm{sgn} (b) M^3 - \lambda M^5 + \nabla^2 M +\theta\left(\vec{r},t\right),
\label{ke2}
\end{align}
where $\gamma = \bar{\gamma}/\sqrt{|a|}$, $\mathrm{sgn}(x)=x/|x|$, and $\lambda = |a|d/|b|^2 >0$. {The dimensionless noise satisfies
\beq
\left\langle \theta(\vec{r'},t')\theta(\vec{r''},t'') \right\rangle = 2\epsilon\delta(\vec{r'}-\vec{r''})
\delta\left(t'-t''\right) ,
\label{fdr}
\eeq
where $\epsilon$ is the noise strength. This is related to temperature \cite{pw09} as 
\beq
\epsilon = \frac{\gamma T |b|}{|a|^{(5-d)/2} K^{d/2}} ,
\eeq
where $d$ is the spatial dimensionality. In principle, the actual forms of $\gamma$ and $\theta$ can be derived from a microscopic field theory using real-time nonequilibrium dynamics \cite{dh98}. This has been attempted for $M^4$-theory \cite{gr94,dr98}, as well as the linear-sigma model coupled to quarks \cite{nl11}.} However, for simplicity, we choose the simple form of Eq.~(\ref{ke2}) which allows a clear distinction of the roles played by the dissipation and inertial terms in the evolution dynamics. The present paper is complementary to Ref.~\cite{ak10}, where we studied ordering dynamics in Eq.~(\ref{ke2}) {\it without} an inertial term.

In this paper, we have presented our results in dimensionless units of space and time. One can obtain the corresponding physical units by multiplying with the appropriate length-scale $\xi$ and time-scale $t_0$. For this, we need to estimate the strength of the interfacial energy $K$. We calculate the surface tension as $\sigma = \sqrt{K}(|a|^{3/2}/|b|)\int dz~(dM_s/dz)^2$, where $M_s(z)$ denotes the 1-dimensional static solution of Eq.~(\ref{ke2}) with $\theta = 0$ \cite{pw09}. For quark matter, $\sigma$ is poorly known and varies from 10-100 MeV/$\text{fm}^2$ at small temperatures \cite{hc93}. {If we take $\sigma \simeq 50$ MeV/$\text{fm}^2$ as in Ref.~\cite{lc92}, at $T=10$ MeV and $\mu = 321.75$ MeV, we  estimate $\xi = \sqrt{K/|a|} \simeq 2.8$ fm and $t_0 = 1/\sqrt{|a|} \simeq 5.1$ fm \cite{ak10, kk92}. On the other hand, recent estimates using effective models like the linear-sigma model \cite{ef10}, Polyakov
quark meson model \cite{rj12} as well as the NJL model \cite{vk12} suggest a lower value of $\sigma \simeq 5$-20 MeV/$\text{fm}^2$. Then the corresponding length scale will also be reduced to, e.g., $\xi \simeq 0.56$ fm for $\sigma=10$ MeV/$\text{fm}^2$.}

We consider a system (in the disordered phase with $M \simeq 0$), which becomes thermodynamically unstable when it is rapidly quenched below the critical lines I or II in Fig.~\ref{fig1}. The subsequent evolution of the system is characterized by the emergence and growth of domains of the ordered phase with non-zero $|M|$. In Sec.~\ref{case1}, we study phase-transition kinetics by quenching through II. The corresponding parameter values are $(a/\Lambda^2,b,d\Lambda^2) = (-1.6 \times 10^{-2}, 9 \times 10^{-2}, 7.1 \times 10^{-2})$ with $\lambda = |a|d/|b|^2 = 0.14$. [This set of Landau parameters corresponds to $(\mu,T)=(231.6, 85)$ MeV \cite{ak10}.] Here the system evolves via spinodal decomposition. In Sec.~\ref{case2}, we consider a quench into the metastable region of the phase diagram. This is achieved by shallow quenching through line I in Fig.~\ref{fig1}, i.e., quenching to a point between I and S$_1$ (marked by a cross in Fig.~\ref{fig1}). Here, the system evolves via nucleation and growth of droplets of the favored phases. This case is studied using the parameter values $(a/\Lambda^2, b, d\Lambda^2) = ( 3.5 \times 10^{-3}, -0.1, 0.4)$ with  $\lambda = 0.14$. [This set of Landau parameters corresponds to $(\mu,T)$=(321.8,10) MeV \cite{ak10}.]

\subsection{Early-time Behavior of the Solution}
\label{STB}

First, we study the early-time behavior of the deterministic version of Eq.~(\ref{ke2}) ($\theta = 0$). We linearize it around an extremum point $\bar{M}$ by replacing $M(\vec{r},t) = \bar{M} + \phi (\vec{r},t)$. In Fourier space, the linearized equation becomes
\begin{align}
\frac{\partial^2}{\partial t^2}\;\phi(\vec{k},t) + \gamma\frac{\partial}{\partial t}\;\phi(\vec{k},t)+ (-\alpha + k^2) \phi(\vec{k},t) = 0,
\label{lke4}
\end{align}
where $\alpha=-f''(\bar{M})$. We have $\alpha > 0$ when $\bar{M}$ is a local maximum, and $\alpha < 0$ when $\bar{M}$ is a local minimum. Equation (\ref{lke4}) is a homogeneous second-order differential equation, and one can write the general solution as
\begin{eqnarray}
\phi(\vec{k},t) &=& A_1 e^{\Lambda_+ (\vec{k})t} + A_2 e^{\Lambda_- (\vec{k})t} , \nonumber \\
\Lambda_{\pm}(\vec{k}) &=& \frac{-\gamma \pm \sqrt{\gamma^2 + 4(\alpha - k^2)}}{2} .
\label{ls1}
\end{eqnarray}
Here $A_1$ and $A_2$ are constants. In the absence of dissipation ($\gamma = 0$), we have
\beq
\Lambda_{\pm} = \pm \sqrt{\alpha - k^2} .
\eeq

First, consider the case $\alpha > 0$. There is an instability for short wavelengths ($k < \sqrt{\alpha}$) with $\Lambda_+ (\vec{k}) > 0$. Thus, there is an exponential growth of fluctuations about a local maximum of the free energy. This is valid even in the limit of no dissipation. For $\alpha < 0$, there is no instability and fluctuations are exponentially damped. The damping is relaxational for $k^2 < (\gamma^2 - 4 |\alpha|)/4$, and oscillatory for $k^2 > (\gamma^2 - 4 |\alpha|)/4$. In the limit of no dissipation, the dynamics is purely oscillatory.

\section{Quench through Second-order Line}
\label{case1}

In this section, we consider deep quenches through the second-order line (II) in Fig.~\ref{fig1}. The chiral transition occurs when we quench from $a>0$ (with $M=0$) to $a<0$, where the free energy has a double-well structure (cf. Fig.~\ref{fig1}). The chirally-symmetric phase is now unstable, and evolves to the stable massive phase via spinodal decomposition. The appropriate form of the evolution equation is 
\begin{align}
\frac{\partial^2 M}{\partial t^2} + \gamma\frac{\partial M}{\partial t} = M - M^3 - \lambda M^5 + \nabla^2 M +\theta\left(\vec{r},t\right).
\label{ke3}
\end{align}
We solve Eq.~(\ref{ke3}) numerically using a simple Euler-discretization scheme with initial velocity $\partial M/\partial t |_{t=0} = 0$. {The initial state of the system is prepared as  $M(\vec{r},0) = 0 \pm \delta M(\vec{r},0)$, where $\delta M$ is a random number uniformly distributed in the range [$-0.25, +0.25$]. This mimics the physical situation where small-amplitude fluctuations are always present. Even if we start with a uniform initial state, thermal noise rapidly generates random fluctuations.}

Our numerical simulations are implemented on a 3-$d$ lattice of size $N^3$ ($N=256$), with periodic boundary conditions in all directions. For all results in this paper, we used the mesh sizes $\Delta x = 1.0$ and $\Delta t = 0.1$, obtained from the linear stability analysis of Eq.~(\ref{ke3}) \cite{ak10,yp87,red88}. {Essentially, we require that the Euler-discretized numerical scheme must respect the stability properties of the homogeneous solutions of Eq.~(\ref{ke3}).} The thermal noise $\theta(\vec{r},t)$ is mimicked by uniformly-distributed random numbers between $[-A_n, A_n]$. {In studies of phase-transition kinetics, it is known that statistical results are unchanged whether we use Gaussian noise or uniformly-distributed noise \cite{yp87,po88}. The appropriate noise amplitude in our simulation is \cite{sp02}
\begin{equation}
A_n = \sqrt{\frac{3\epsilon}{(\Delta x)^d \Delta t}}.
\label{na}
\end{equation}
The results reported here correspond to $\epsilon=0.008$, i.e., $A_n=0.5$. All statistical quantities are obtained as averages over 10 independent runs.}

\begin{figure}[!htb]
\centering
\includegraphics[width=0.95\textwidth]{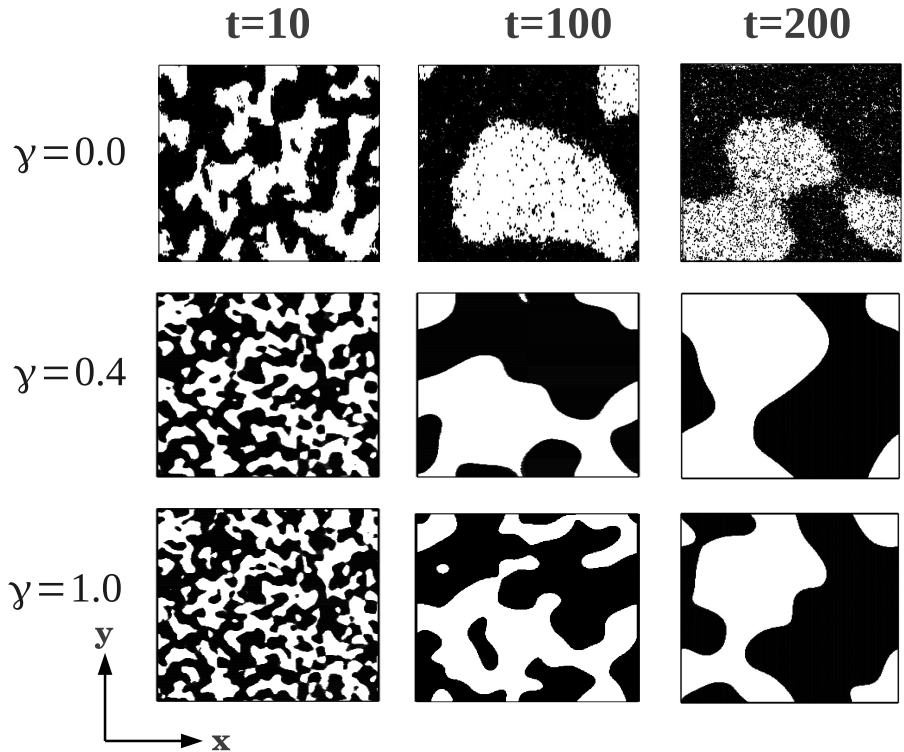}
\caption{Domain evolution of the preferred massive phase: $M=M_+$ (marked black), after a deep temperature quench through the second-order line (II) in Fig.~\ref{fig1}. We show evolution pictures at $t=10, 100, 200$ for three different values of $\gamma$. The frames are the cross-sections at $z=N/2$ of the 3-$d$ snapshots obtained by numerically solving Eq.~(\ref{ke3}) with $\lambda=0.14$. The noise strength is $\epsilon=0.008$.}
\label{fig2} 
\end{figure}

In Fig.~\ref{fig2} we show the evolution of Eq.~(\ref{ke3}) from a disordered initial state. To study the effect of dissipation, we chose $\gamma=0, 0.4, 1.0$. After the quench, the system rapidly evolves into domains of the massive phase with $M\simeq M_+$ (marked black) and $M\simeq -M_+$ (unmarked). The snapshots show the evolution of the preferred phases at $t=10$, $100$ and $200$. The frames are the cross-sections of the 3-$d$ snapshots at $z=N/2$. {For $\gamma=0$, the dissipative term is absent and we observed a rapid growth of domains of the massive phases (e.g., see the pattern at $t=10$). As expected, the dissipation coefficient $\gamma$ controls the rapid growth achieved due to the inertial term in Eq.~(\ref{ke3}). After the initial rapid growth, domain walls get fuzzier, and domains become less distinctive due to the oscillatory behavior of the system (e.g., see the patterns at $t=100$ and $t=200$). We have also tracked the order-parameter value at a few spatial points in the $\gamma = 0$ case. We observe the occurrence of flips from $\pm M_+ \rightarrow \mp M_+$ on extended time-scales. In spite of these, the domain morphology continues to coarsen as these oscillations are cooperative.}

The system is characterized by a single length scale $L(t)$ as the pattern morphology does not change in time apart from a scale factor. {The morphology is quantitatively studied using the \emph{correlation function} \cite{pw09}:
\beq
C\left(\vec{r},t\right) = \frac{1}{V} \int d\vec{R}\left[ \left\langle M(\vec{R},t)M(\vec{R}+\vec{r},t)\right\rangle - \left\langle M(\vec{R},t)\right\rangle \left\langle M(\vec{R}+\vec{r},t)\right\rangle\right]. 
\label{cf} 
\eeq
Here, $V$ denotes the volume of the system, and the angular brackets denote an average over independent runs. The evolution morphologies are isotropic, so we compute the spherically-averaged correlation function $C(r,t)$ with $r=|\vec r|$.} The existence of the characteristic scale results in a \emph{dynamical scaling} of the correlation function:
{
\beq
C(r,t)=g\left[r/L(t)\right] .
\eeq
}
While  microscopic techniques can be used to measure $C(\vec{r},t)$, scattering experiments probe its Fourier transform, called the \emph{structure factor}:
{
\beq
S(\vec{k},t) = \int d\vec{r}~e^{i\vec{k}\cdot\vec{r}}C\left(\vec{r},t\right).
\label{sf}
\eeq
Again, a spherical average is taken since the system is isotropic.} The structure factor also has a dynamical-scaling form: {
\beq
S(k,t) = L(t)^d f\left[k L(t)\right] ,
\eeq
}
where $f(p)$ is the scaling function \cite{aj94,pw09}.

\begin{figure}[!htb]
\centering
\includegraphics[width=0.6\textwidth]{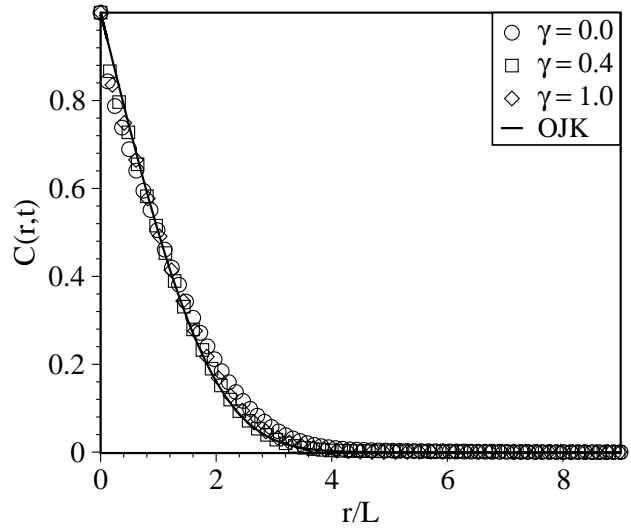}
\caption{Plot of the scaled correlation function, $C(r,t)$ vs. $r/L$, for $\gamma = 0, 0.4, 1.0$ at $t=20$. The length scale $L(t)$ is defined as the distance over which the correlation function decays to half its maximum value [$C(r,t)=1$ at $r =0$]. The solid line denotes the OJK function in Eq.~(\ref{ojk}).} 
\label{fig3}
\end{figure}

We have confirmed numerically (not shown here) that the correlation functions at different times obey dynamical scaling for different $\gamma$-values. In Fig.~\ref{fig3}, we plot the scaled correlation function, $C(r,t)$ vs. $r/L$, for $\gamma = 0, 0.4, 1.0$ at $t=20$. The length scale $L(t)$ is defined as the distance over which the correlation function decays to half its maximum value [$C(r,t)=1$ at $r =0$]. Notice that the scaling functions are numerically indistinguishable showing that the evolution morphologies are the same for different values of $\gamma$. The solid line denotes the Ohta-Jasnow-Kawasaki (OJK) function \cite{ojk82,bp91}:
\beq
g_{\rm OJK} (x) = \frac{2}{\pi} \sin^{-1} \left( e^{-x^2/2} \right) ,
\label{ojk}
\eeq
which characterizes ordering dynamics for the $M^4$-potential in the overdamped limit, i.e., without inertial terms. Clearly, our numerical data is well-described by the OJK function.

\begin{figure}[!htb]
\centering
\includegraphics[width=0.6\textwidth]{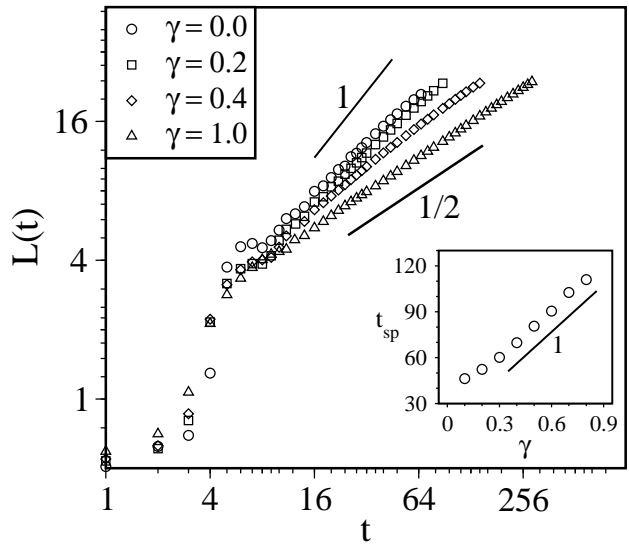}
\caption{Time-dependence of domain size, $L(t)$ vs. $t$, for the evolution depicted in Fig.~\ref{fig2}. There is a crossover from an early-time inertial growth [$L(t) \sim t (\ln t)^{1/2}$] to a late-time Cahn-Allen (CA) growth [$L(t) \sim t^{1/2}$]. The inset plots $t_{\rm sp}$ vs. $\gamma$, where $t_{\rm sp}$ is the time-scale for amplification and saturation of initial fluctuations. {The statistical data shown in this figure is obtained as an average over 10 independent runs.}} 
\label{fig4}
\end{figure}

In Fig.~\ref{fig4}, we plot $L(t)$ vs. $t$ on a log-log scale. {The growth proceeds by the amplification of initial fluctuations, their saturation by the nonlinearity, and subsequent domain coarsening. We denote $t_{\rm sp}$ as the time-scale on which coarsening is initiated, i.e., the time-scale for amplification and saturation of initial fluctuations about $M=0$. This time-scale can be estimated from linear stability analysis (described in Sec.~\ref{STB}) as $t_{\rm sp} \sim 2/(2-\gamma) \simeq 1 + \gamma/2$ for small $\gamma$-values. The inset of Fig.~\ref{fig4} plots $t_{\rm sp}$ vs. $\gamma$.}

{
At this stage, it is relevant to ask how the inertial term affects the growth kinetics. To understand this, we consider the deterministic version ($\theta = 0$) of Eq.~(\ref{ke3}), which we rewrite as
\beq
\frac{\partial^2 M}{\partial t^2} + \gamma\frac{\partial M}{\partial t} = -f'(M) + \nabla^2 M ,
\label{ke3d}
\eeq
where 
\beq
f(M) = -\frac{M^2}{2} + \frac{M^4}{4} + \lambda \frac{M^6}{6} .
\label{poten}
\eeq
The 1-dimensional static (kink) solution $M_s(z)$ of Eq.~(\ref{ke3d}) is the same in the inertial and overdamped cases and obeys
\beq
-f'(M_s) + \frac{d^2 M_s}{dz^2} = 0 .
\label{ms}
\eeq
Equation~(\ref{ms}) gives rise to a tanh-profile (sigmoidal) between $M=-1$ and $M=+1$ for the usual $M^4$-potential: $f(M) = -M^2/2 + M^4/4$. The corresponding kink profile for the potential in Eq.~(\ref{poten}) connects the two vacuum states: $+M_+$ and $-M_+$, where $M_+^2 = (-1+ \sqrt{1+4 \lambda})/(2\lambda)$.}

{
For Eqs.~(\ref{ke3d})-(\ref{poten}), we consider a droplet of $M=+M_+$ shrinking in a background with $M=-M_+$. If the radius of the droplet is $R(t)$, then
\beq
M(\vec{r},t) \simeq h[r-R(t)] \equiv h(\eta) ,
\label{sig}
\eeq
where $h(\eta)$ is a sigmoidal profile whose derivative is sharply peaked at $r=R(t)$. Replacing Eq.~(\ref{sig}) in Eq.~(\ref{ke3d}), we obtain
\beq
h'' \left(\frac{dR}{dt}\right)^2 - h' \frac{d^2 R}{dt^2} - \gamma h' \frac{d R}{dt} = h'' + \frac{d-1}{r} h' - f'(h) ,
\label{hpp1}
\eeq
or
\beq
0 = h'' \left[ 1 - \left(\frac{dR}{dt}\right)^2 \right] + h' \left( \frac{d-1}{r} + \gamma \frac{d R}{dt}
+ \frac{d^2 R}{dt^2} \right) - f'(h) .
\label{hpp2}
\eeq
We multiply Eq.~(\ref{hpp2}) by $h'$ and integrate through the interface. The first term on the RHS drops out because $h' = 0$ as $\eta \rightarrow \pm \infty$, and the third term drops out because $f(M_+) = f(-M_+)$. This yields the kinetic equation for droplet shrinkage:
\beq
\frac{d^2 R}{dt^2} + \gamma \frac{d R}{dt} = - \frac{d-1}{R} .
\eeq
}

{The analogous growth equation for the domain scale $L(t)$ is \cite{aj94,pw09}
\beq
\frac{d^2 L}{dt^2} + \gamma \frac{d L}{dt} = \frac{\sigma}{L} ,
\eeq
where the RHS is identified as the curvature for a domain of size $L$. At short times ($t \ll t_c$), the growth law is determined by the inertial term as \cite{bo84}
\beq
L(t) \sim \sqrt{\sigma} t \left[ \ln (\sqrt{\sigma} t) \right]^{1/2} .
\eeq
The long-time ($t \gg t_c$) kinetics is determined by the dissipative term as
\beq
L(t) \sim \left( \frac{\sigma t}{\gamma} \right)^{1/2} ,
\eeq
which is the usual Cahn-Allen (CA) growth law. The crossover-time scales as $t_c \sim \gamma^{-1}$. In Fig.~\ref{fig4}, we have plotted straight lines corresponding to $L(t) \sim t$ and $L(t) \sim t^{1/2}$, the two limiting behaviors of the growth law.}

\section{Quench through First-order Line}
\label{case2}

Let us next focus on the ordering dynamics for shallow quenches through the first-order line ($b < 0$) in Fig.~\ref{fig1}. For $b < 0$, the first-order chiral transition occurs at $a<a_c=3|b|^2/(16d)$. We consider quenches from the disordered state (with $M=0$) at $a>a_c$ to $a<a_c$. If we quench to $a<0$, the free energy has a double-well structure, as shown in Fig.~\ref{fig1}. In this case, the ordering dynamics is analogous to the previous case with $b>0$, discussed in Sec.~\ref{case1}. We have numerically confirmed that the domain growth scenario is similar to that shown in Figs.~\ref{fig2}--\ref{fig4}. Subsequently, we focus only on quenches to $0<a<a_c$. The appropriate form of the dimensionless time-dependent evolution equation is [cf. Eq.~(\ref{ke2})]
\begin{align}
\frac{\partial^2 M}{\partial t^2} + \gamma\frac{\partial M}{\partial t} = -M + M^3 - \lambda M^5 + \nabla^2 M +\theta\left(\vec{r},t\right).
\label{ke4}
\end{align}
The free-energy extrema of the corresponding potential,
\begin{eqnarray}
f\left(M \right) = \dfrac{M^2}{2} - \dfrac{M^4}{4} + \lambda \dfrac{M^6}{6} ,
\label{tp6}
\end{eqnarray}
are located at $M=0$, $\pm M_+$, and $\pm M_-$, where $M_{+} = \left[(1+\sqrt{1-4\lambda})/(2\lambda) \right]^{1/2}$ and $M_{-} = \left[(1-\sqrt{1-4\lambda})/(2\lambda) \right]^{1/2}$. The extrema $M=0, \pm M_+$ are the local minima with $f(\pm M_+)<f(0)=0$ for $\lambda<\lambda_c = |a_c|d/|b|^2 = 3/16$.

\subsection{Phase Plane Analysis and Bubble Growth}
\label{PPA}

Before we study the ordering dynamics of Eq.~(\ref{ke4}), it is instructive to understand the nature of the traveling-wave solutions. Our analytical understanding of the domain growth problem is based on the dynamics of interfaces (kinks and anti-kinks) that separate regions enriched in the two states $+M_+$ and $-M_+$. For the case with $b>0$ and $a<0$, discussed in Sec.~\ref{case1}, the kinks have tanh-profiles with small corrections due to the $M^6$-term in the potential [cf. Eq.~(\ref{ms})].

\begin{figure}[!htb]
\centering
\begin{tabular}{c }
\includegraphics[width=0.65\textwidth]{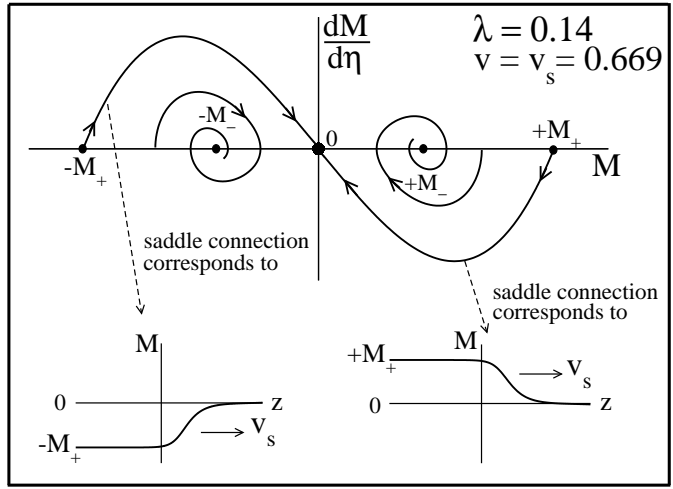}
\end{tabular}
\caption{Phase portrait of the dynamical system in Eqs.~(\ref{ode1}) for $\lambda = 0.14$, $\gamma = 0.5$. The phase portrait is plotted for $v_s = 0.669$, where $v_s$ corresponds to the appearance of saddle connections from $-M_+ \rightarrow 0$ and $+M_+ \rightarrow 0$. These correspond to kinks traveling with velocity $v_s > 0$.} 
\label{fig5}
\end{figure}

We consider the deterministic version of Eq.~(\ref{ke4}) in $d=1$:
\begin{equation}
\frac{\partial^2 M}{\partial t^2} + \gamma\frac{\partial M}{\partial t} = -M + M^3-\lambda M^5 + \frac{\partial^2 M}{\partial z^2}.
\label{dke4}
\end{equation}
We focus on traveling-wave solutions of this equation, $M\left(z,t\right) \equiv M\left(z-vt\right) \equiv M\left(\eta\right)$ with velocity $v>0$. This reduces Eq.~(\ref{dke4}) to an ordinary differential equation:
\begin{align}
\left(1-v^2\right)\frac{d^2 M}{d\eta^2} + \gamma v\frac{d M}{d\eta} - M + M^3 -\lambda M^5 = 0.
\label{ode}
\end{align}
Equation~(\ref{ode}) is equivalent to a 2-$d$ dynamical system:
\begin{align}
\frac{d M}{d\eta} =& \; y, \notag \\
\left(1-v^2\right)\frac{d y}{d\eta} =& \; M - M^3 + \lambda M^5 -\gamma vy.
\label{ode1}
\end{align}
The phase portrait of this system will enable us to identify traveling-wave solutions of Eq.~(\ref{dke4}). The relevant fixed points (FPs) of Eqs.~(\ref{ode1}) are $(M,y)= (0,0)$, $(\pm M_-,0)$, $(\pm M_+,0)$ \cite{ak10}. In Fig.~\ref{fig5}, we show the phase portraits for $\lambda=0.14$ ($<\lambda_c = 0.1875$), $\gamma = 0.5$, and $v = v_s$, where $v_s$ corresponds to the appearance of the saddle connections from $-M_+ \rightarrow 0$ and  $+M_+ \rightarrow 0$. These correspond to kinks traveling with velocity $v_s >0$, as shown in Fig.~\ref{fig5}. Here, our analysis has been for the case with $v>0$, but it is straightforward to extend it to the case with $v<0$. In the latter case, the portrait in Fig.~\ref{fig5} is inverted, and the saddle connections (kinks) are from $0 \rightarrow -M_+$ and  $0 \rightarrow +M_+$.

\begin{figure}[!htb]
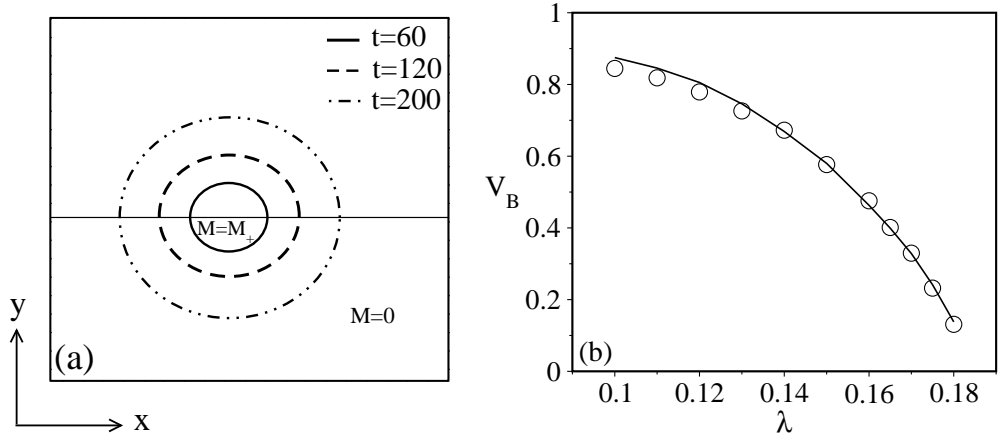

\centering
\begin{tabular}{c c }
\includegraphics[width=0.43\textwidth]{f6a.eps} &
\includegraphics[width=0.50\textwidth]{f6b.eps}\\
\end{tabular}
\caption{(a) Growth of a droplet of the preferred phase ($M = M_+$) in a background of the metastable phase ($M=0$) for $\lambda = 0.14$. We show the boundary of the droplet at three different times, as specified. (b) Plot of the bubble growth velocity $v_B$ vs. $\lambda$. The circles denote the numerical data, while the solid line corresponds to the result from a phase-plane analysis.}
\label{fig6} 
\end{figure}

Next, we study the growth dynamics of a bubble or droplet. For the free energy considered, the first-order transition proceeds via the initial metastable phase breaking into droplets of the preferred phases. In Fig.~\ref{fig6}(a), we show the growth of a droplet of the preferred massive phase ($M=+M_+$) in the background of the metastable phase ($M=0$). {We obtain this evolution by solving Eq.~(\ref{ke4}) in $d=2$ with $\gamma=0.5$, $\lambda=0.14$,  and $\theta=0$.
We start with an initial configuration of a bubble of radius $R_0 > R_c$ such that $M(r)=M_+$ for $r<R_0$, and $M(r)=0$ for $r>R_0$. The slope of $R(t)-R_0$ vs. $t$ gives the bubble velocity $v_B$ -
 In Fig.~\ref{fig6}(b), we plot bubble velocity $v_B$ vs. $\lambda$.
 Here, we have taken $R_0=10$ as the initial size of the droplet 
($R_0 > R_c \simeq 8$, which is the critical size for these parameter values).} Our numerical data is in good agreement with $v_s(\lambda)$, which is obtained from the phase-plane analysis.

\subsection{Chiral Transition Kinetics}
\label{ctk}

Next, we consider the ordering dynamics of Eq.~(\ref{ke4}) from a disordered state. We use the same numerical scheme described in Sec.~\ref{case1}. The initial state with massless quarks ($M=0$) is now a metastable state of the potential, {and phase separation proceeds via nucleation and growth of droplets of the preferred phase ($M = \pm M_+$).} Therefore, the thermal noise $\theta(\vec{r},t)$ must be sufficiently large to enable the system to escape from the metastable state on a reasonable time-scale: a suitable value for $\lambda = 0.14$ is $\epsilon = 0.6$. However, the asymptotic behavior of domain growth in both the unstable and metastable cases is insensitive to the noise term \cite{po88}.

\begin{figure}[!ht]
\centering
\includegraphics[width=0.95\textwidth]{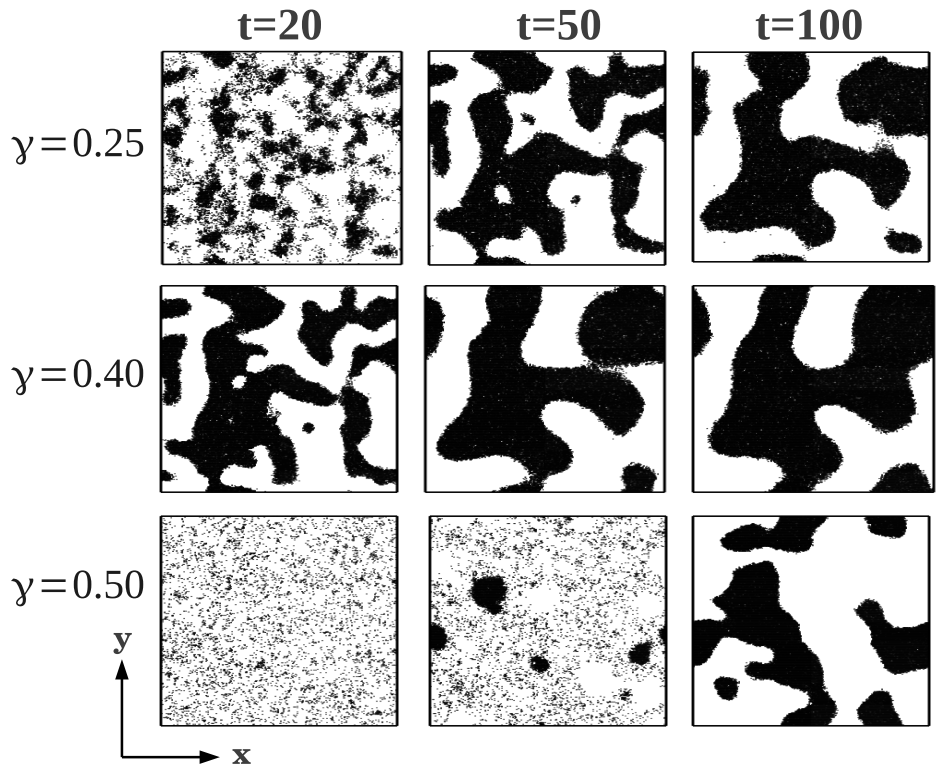}
\caption{Domain growth after a shallow temperature quench through the first-order line (I) in Fig.~\ref{fig1} for $\gamma=0.25, 0.4, 0.5$. The frames show the evolution of the preferred phase with $M=+M_+$ (marked black) at times $t=20$, $50$ and $100$, respectively. Nucleation is fastest for moderate values of $\gamma$, as explained in the text.} 
\label{fig7}
\end{figure}

In Fig.~\ref{fig7}, we show the domain growth of the system for $\gamma = 0.25$, $0.4$ and $0.5$. The frames are cross-section of the 3-$d$ snapshots at $z=N/2$, and show the evolution of one of the preferred phases ($M=M_+$) at times $t=20$, $50$ and $100$. Typically, the evolution of the system begins with the nucleation of droplets in the early stages: {droplets larger than a critical size $R_c$ (supercritical) grow, whereas those with $R<R_c$ (subcritical) shrink. In nucleation theory \cite{pw09}, $R_c$ is defined by the balance between {\it free-energy reduction} due to the appearance of the bulk droplet and {\it free-energy increase} due to the surface tension at the droplet boundary. In the present simulation, the critical radius of the bubble $R_c \simeq 8$ in dimensionless units. If we convert this into physical units, $R_c \simeq 4.5$ fm
($\xi$=0.56 fm for $\sigma$=10MeV/fm$^2$).  This value may be compared
with the critical bubble size estimated as $2\sigma/\Delta p$ within
thin wall approximation \cite{fj10}, where $\Delta p$ is the 
pressure difference between the metastable and stable states. In the present 
model with the parmeters considered, $\Delta p \simeq 3.4$ MeV/fm$^3$. 
If we take the surface tension to be 10 MeV/fm$^2$, then the critical bubble size is about 6 fm.}

The droplets grow very rapidly and fuse to form bi-continuous domain structures, a characteristic of late-stage domain growth. The effect of dissipation on nucleation and growth can be understood by comparing the evolution patterns at different $\gamma$-values in Fig.~\ref{fig7}. We observed that the system takes more time to nucleate for the limiting $\gamma$-values (i.e.,$\gamma \rightarrow 0$ and $\gamma \rightarrow \infty$). However, for intermediate values ($\gamma \simeq 0.4$), it takes less time. {To understand this behavior, we consider {\it Kramer's escape problem} for a barrier, as discussed by Hanggi \cite{ph86}. Hanggi studies the homogeneous version of Eq.~(\ref{ke4}), and the corresponding crossover time from $M=0$ (the metastable state) to $M=M_+$ (the stable state). This crossover time is proportional to the nucleation time $t_n$ in our domain growth problem. We designate $\omega_b$ as the natural vibration frequency about the barrier location ($M_-$). For moderate to large dissipation ($\gamma \gg \omega_b$), the nucleation time
\beq
t_n \sim \left(\sqrt{\frac{\gamma^2}{4} + \omega_b^2} - \frac{\gamma}{2} \right)^{-1} ,
\eeq
so that $t_n \sim \gamma$ as $\gamma \rightarrow \infty$. For small dissipation ($\gamma \ll \omega_b$), we have
\beq
t_n \sim \frac{1}{\gamma} ,
\eeq
so that $t_n \rightarrow \infty$ as $\gamma \rightarrow 0$.}

\begin{figure}[!htb]
\centering
\includegraphics[width=0.6\textwidth]{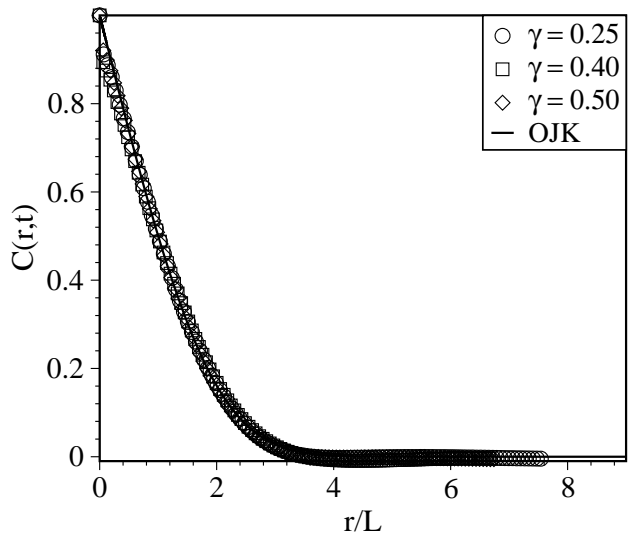}
\caption{Analogous to Fig.~\ref{fig3}, but for the evolution depicted in Fig.~\ref{fig7}.} 
\label{fig8}
\end{figure}

In Fig.~\ref{fig8}, we plot the scaled correlation function for the evolution depicted in Fig.~\ref{fig7}. All the data sets correspond to $t=100$. Again, we observe that the scaling functions are universal for the late-stage dynamics, subsequent to the nucleation regime. Thus, in all cases considered here, the domain growth morphology is well-described by the OJK function in the asymptotic regime.

\begin{figure}[!htb]
\centering
\includegraphics[width=0.85\textwidth]{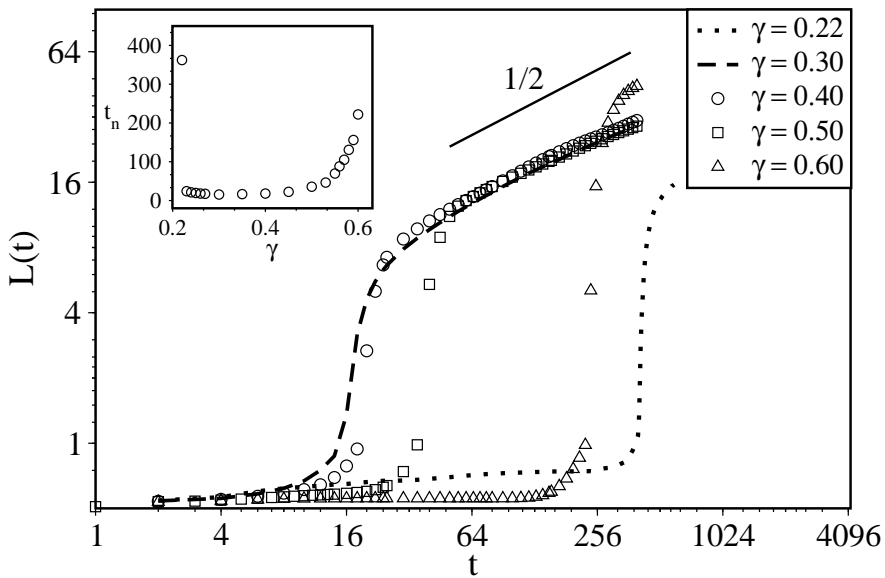}
\caption{Time-dependence of the domain size, $L(t)$ vs. $t$, for different $\gamma$-values. There is no growth in the early stages when droplets are being nucleated. The asymptotic growth is consistent with the CA growth law, $L(t) \sim t^{1/2}$. The inset shows the $\gamma$-dependence of the nucleation time $t_n$ for the onset of domain growth.}
\label{fig9} 
\end{figure}

In Fig.~\ref{fig9}, we plot the domain size [$L(t)$ vs. $t$] on a log-log scale. We observe that the growth process begins once the nucleation of droplets is over. {The onset time for domain growth is the nucleation time $t_n$, which is shown in the inset of Fig.~\ref{fig9} for different values of $\gamma$. As expected from the earlier discussion, $t_n \rightarrow \infty$ as $\gamma \rightarrow 0$ or $\gamma \rightarrow \infty$. The intermediate and asymptotic growth regimes are analogous to those described for Fig.~\ref{fig4}, i.e., crossover from $L(t) \sim t (\ln t)^{1/2}$ to $L(t) \sim t^{1/2}$. In Fig.~\ref{fig9}, we have focused on the $\gamma$-dependence of $t_n$, rather than the asymptotic growth laws.}

\section{Summary and Discussion}
\label{summary}

To conclude this paper, we summarize and discuss the results presented here. We have studied the kinetics of chiral phase transitions in {strong interactions within the ambit of the Nambu-Jona-Lasinio (NJL) model as an effective model to study chiral symmetry breaking and it restoration in QCD. 
}. First, we have proposed a quantitative mapping between the free energy of the NJL model, and the $M^6$-Landau potential. This mapping enables us to identify the relevant time-scales and length-scales within the model kinetics. Near the phase transition, we can relate ($\mu,T$) to the coefficients of the Landau model. However, we have considered parameter values far from the critical points, where it is more appropriate to interpret the Landau coefficients as phenomenological quantities.

We have studied the kinetics of transitions from the massless quark (disordered) phase to the massive quark (ordered) phase, resulting from a sudden quench in the {system parameters}. We model the kinetics via a time-dependent Ginzburg-Landau (TDGL) equation with both dissipative and inertial terms. The inertial term arises naturally in the context of strong interaction kinetics, though it is usually neglected in condensed matter applications. We are particularly interested in the effect of the inertial term on phase-transition kinetics.

For deep quenches, the massless phase is spontaneously unstable and evolves via {\it spinodal decomposition}. In the purely inertial case ($\gamma=0$), we expect a rapid spinodal decomposition with growth law $L(t) \sim t (\ln t)^{1/2}$. In this limit, the order parameter value does not relax to an equilibrium value due to the oscillatory dynamics of the system. In the presence of dissipation ($\gamma > 0$), there is a crossover at $t_c \sim \gamma^{-1}$ from the faster growth regime to Cahn-Allen (CA) growth with $L(t) \sim t^{1/2}$. Further, the evolution morphologies show self-similar scaling, and can be quantitatively characterized by the order-parameter correlation function or its Fourier transform, the structure factor.

For shallow quenches through the I-order line, the massless phase is metastable. The system evolves via nucleation and the growth of droplets of the preferred phase. The dissipation factor $\gamma$ affects the duration of the nucleation regime: the nucleation time $t_n \rightarrow \infty$ as $\gamma \rightarrow 0, \infty$. Once the nucleation regime is over, droplets quickly merge to form  bi-continuous spatial domains {and again the massive phase grows as $L(t) \sim t^{1/2}$ for long times.}

{
Before concluding, we should discuss the relevance of these results for QCD phenomenology and experiments. In the context of heavy-ion collisions, given the uncertain values of dimensional quantities for quark matter (e.g., surface tension, dissipation), it is not clear whether the system equilibrates completely within the life-time of the fireball. If the system is almost equilibrated, the features of the coarsening morphology are similar for quenches through both first- and second-order lines in the phase diagram. However, if the equilibration time-scale is much larger than the fireball life-time, the morphology is very different for quenches through the first-order line, with the system evolving through nucleation of droplets. These signatures of a first-order transition are experimentally relevant because they imply the existence of a critical end point (CEP) in the QCD phase diagram. As a matter of fact, experimental studies of such signatures may be more convenient than directly searching for the CEP via critical fluctuations. To date, the latter approach has not provided conclusive evidence of the existence of a CEP, presumably due to the smallness of the critical region.}

\vspace{0.5cm}
\noindent {\bf Acknowledgements} \\
AS thanks CSIR (India) for providing financial support. HM would like to thank the School of Physical Sciences, Jawaharlal Nehru University for hospitality. {The authors are also very grateful to an annonymous referee for constructive and helpful comments.}

\newpage

\end{document}